\documentclass[journal=nalefd,manuscript=letter]{achemso}

\usepackage[version=3]{mhchem}

\author{Tao Jia}
\affiliation{Stanford Institute for Materials and Energy Sciences, SLAC National Accelerator Laboratory, Menlo Park, California 94025, USA}
\alsoaffiliation{Geballe Laboratory for Advanced Materials, Departments of Physics and Applied Physics,
Stanford University, Stanford, California 94305, USA}
\altaffiliation{These two authors contributed equally to this work.}

\author{Slavko N. Rebec}
\affiliation{Stanford Institute for Materials and Energy Sciences, SLAC National Accelerator Laboratory, Menlo Park, California 94025, USA}
\alsoaffiliation{Geballe Laboratory for Advanced Materials, Departments of Physics and Applied Physics,
Stanford University, Stanford, California 94305, USA}
\altaffiliation{These two authors contributed equally to this work.}

\author{Shujie Tang}
\affiliation{Stanford Institute for Materials and Energy Sciences, SLAC National Accelerator Laboratory, Menlo Park, California 94025, USA}
\affiliation{Geballe Laboratory for Advanced Materials, Departments of Physics and Applied Physics,
Stanford University, Stanford, California 94305, USA}

\author{Kejun Xu}
\affiliation{Geballe Laboratory for Advanced Materials, Departments of Physics and Applied Physics,
Stanford University, Stanford, California 94305, USA}

\author{Hafiz M. Sohail}
\affiliation{Stanford Institute for Materials and Energy Sciences, SLAC National Accelerator Laboratory, Menlo Park, California 94025, USA}
\alsoaffiliation{Geballe Laboratory for Advanced Materials, Departments of Physics and Applied Physics,
Stanford University, Stanford, California 94305, USA}

\author{Makoto Hashimoto}
\affiliation{Stanford Synchrotron Radiation Lightsource, SLAC National Accelerator Laboratory, Menlo Park, California 94025, USA}

\author{Dong-Hui Lu}
\affiliation{Stanford Synchrotron Radiation Lightsource, SLAC National Accelerator Laboratory, Menlo Park, California 94025, USA}

\author{Robert G. Moore}
\affiliation{Stanford Institute for Materials and Energy Sciences, SLAC National Accelerator Laboratory, Menlo Park, California 94025, USA}

\author{Zhi-Xun Shen}

\affiliation{Stanford Institute for Materials and Energy Sciences, SLAC National Accelerator Laboratory, Menlo Park, California 94025, USA}
\alsoaffiliation{Geballe Laboratory for Advanced Materials, Departments of Physics and Applied Physics,
Stanford University, Stanford, California 94305, USA}
\email{zxshen@stanford.edu}
\title
  {Suppression of Charge Density Wave by Substrate Induced Doping on TiSe$_2$/TiO$_2$ Heterostructure}


\begin{document}

% \begin{tocentry}

% \includegraphics[]{ToTv2.eps}

% \end{tocentry}

\begin{abstract}
Substrate engineering provides an opportunity to modulate the physical properties of quantum materials in thin film form.
Here we report that TiSe$_2$ thin films grown on TiO$_2$ have unexpectedly large electron doping that suppresses the charge density wave (CDW) order.
This is dramatically different from either bulk single crystal TiSe$_2$ or TiSe$_2$ thin films on graphene.
The epitaxial TiSe$_2$ thin films can be prepared on TiO$_2$ via molecular beam epitaxy (MBE) in two ways: by conventional co-deposition using selenium and titanium sources, and by evaporating only selenium on reconstructed TiO$_2$ surfaces.
Both growth methods yield atomically flat thin films with similar physical properties.
The electron doping and subsequent suppression of CDW order can be explained by selenium vacancies in the TiSe$_2$ film, which naturally occur when TiO$_2$ substrates are used. 
This is due to the stronger interfacial bonding that changes the ideal growth conditions.
Our finding provides a way to tune the chemical potential of chalcogenide thin films via substrate selection and engineering.

\end{abstract}

\section{Introduction}
Thin film growth using MBE has been shown to be a powerful tool for studies of correlated materials\cite{LAOSTO,WangXue2012cpl,Logvenov09Science}.
Physical properties of correlated thin films can be significantly changed through substrate selection and engineering.
For instance, titanate substrates including SrTiO$_3$, anatase and rutile TiO$_2$ have been found to have multiple effects on FeSe films, including large and anisotropic strain, charge transfer, interfacial electron-phonon coupling and superconducting transition temperature enhancement\cite{WangXue2012cpl,Lee14Nature,DingXue16prl,RJ17prl}.
It is therefore interesting to test the effect of titanate substrates on other thin films of correlated electron systems. 

Layered transition-metal dichalchogenides (TMDC) have been extensively studied due to their interesting physical properties, including the emergence of superconductivity and CDW order\cite{Rossnagel11,Manzeli17}.
Specifically, TiSe$_2$ has attracted significant interest due to the controversy over the driving force of its CDW order.
TiSe$_2$ bulk single crystals undergo a phase transition at around 200 K into a CDW state characterized by the 3D commensurate ($2\times2\times2$) wave vector \cite{Salvo76}.
There are multiple explanations of the origin of the CDW in TiSe$_2$, including band Jahn-Teller effect\cite{Hughes77}, exciton condensation \cite{Wilson77SSC,Cercellier2007} and conventional Fermi-surface nesting\cite{Suzuki84SSC}.
Monolayer thin films of TiSe$_2$ have been grown on bilayer graphene using MBE, yielding similar band structure, and CDW with a 2D commensurate ($2\times2$) wave vector. The CDW transition temperature for monolayer TiSe$_2$ on graphene is slightly higher than that of bulk crystals \cite{ChenChiang15,SugawaraTakahashi16}.
In multilayer thin films of TiSe$_2$ on graphene, the CDW wave vector changes from 2D-like to 3D-like and the transition temperature decreases to bulk value as thickness increases\cite{ChenChiang16}.
The intriguing physical properties of TiSe$_2$ make it an attractive testbed for the study of substrate effects on thin films.

In this work, we grew TiSe$_2$ on rutile TiO$_2$ (100) (TiSe$_2$/TiO$_2$) and bi-layer graphene (TiSe$_2$/graphene) substrates using MBE and characterized its electronic structure via angle-resolved photoemission spectroscopy (ARPES).
We find that the TiSe$_2$ films on TiO$_2$ (TiSe$_2$/TiO$_2$) are heavily electron doped and do not show the signatures of a CDW in ARPES spectra.
Furthermore, after depositing selenium on vacuum-annealed TiO$_2$ substrates (thereafter noted as Se:TiO$_2$), Se atoms reacts with the reconstructed substrate surface, and organizes into epitaxial films of TiSe$_2$.
Our experiments on TiSe$_2$/TiO$_2$ films provide a way to tune the chemical potential and change the ground state property of thin films through substrate engineering.

\section{Electronic Band Structure}
Previous researchers have reported the band structure of TiSe$_2$ (either in bulk form or TiSe$_2$/graphene)\cite{Rossnagel02prb, ChenChiang15}, and the results of TiSe$_2$ bulk single crystals and TiSe$_2$/graphene are very similar.
Near the Fermi level ($E_F$), there is a pair of bands centered at the $\Gamma$ point.
At the M point, the bottom of an electron band reaches just below $E_F$.
We measured the Fermi surface and band structure of TiSe$_2$/graphene at $\sim$ 20 K, shown in Fig 1(b, c).
At this measurement temperature, TiSe$_2$/graphene is in the CDW phase and the ARPES data clearly show that the hole bands at $\Gamma$ are folded to M. There is little signal of the folding from the electron band at M to $\Gamma$, which is consistent with previous ARPES works\cite{Rossnagel02prb, ChenChiang15}.
From the Fermi surface map we observe six tiny pockets formed by electron bands at M.
These results are consistent with the previous research on TiSe$_2$/graphene\cite{ChenChiang15,SugawaraTakahashi16,ChenChiang16}.

The Fermi surface and band structure of TiSe$_2$/TiO$_2$ at the same measurement temperature of 20 K, shown in Fig 1(e-f), exhibit a few stark differences from those of TiSe$_2$/graphene.
The bands are no longer folded, indicating the lack of CDW order.
Also the electron pockets are much larger for TiSe$_2$/TiO$_2$, representing a doping of $n\approx0.16(2)$ electron per unit cell, calculated from the Fermi surface volume.
In contrast, TiSe$_2$/graphene has a doping of merely $n = 0.02$ over the perfectly stoichiometric compound (see Fig 1(b)). The doping of TiSe$_2$/graphene is nonzero possibly due to Se vacancies.
Besides those two features, below the main electron band near the M point (denoted as $\alpha$ in Fig. 1(f)) there is a weaker copy of the band around 100 meV below the main band and crossing the Fermi level, denoted as $\alpha'$ band.
This ``copy'' band is more clearly seen in the second energy derivative of the ARPES spectra in Supplemental Figure 2.
This cannot be attributed to the quantum well effects in the finite-thickness films, as the energy difference between $\alpha$ and $\alpha'$ is thickness independent up to 24 monolayers (ML) (see Supplemental Fig. 1).
Further work needs to be done to explain the origin of the band $\alpha'$.

All the three features mentioned above, including lack of CDW, large electron doping and the ``copy" band $\alpha'$ at M, have little thickness dependence from 3ML to 24ML, as is demonstrated in Supplemental Fig. 1.

Rutile TiO$_2$ (100) surface undergoes a (3 $\times$ 1) reconstruction above 670 $^\circ$C. During this process parts of surface atoms are lost, leading to a terraced surface with exposed atoms of titanium \cite{DieboldTiO2}.
Surprisingly, after annealing the rutile TiO$_2$ substrate above the surface reconstruction temperature in vacuum, we can grow epitaxial thin films of TiSe$_2$ by only depositing Se.
Given that, we speculate that Se atoms react with the Ti atoms exposed to vacuum due to the reconstruction, and form TiSe$_2$ (denoted before as Se:TiO$_2$).
The electronic band structure of Se:TiO$_2$, shown in Fig 1(i), is very similar to that of TiSe$_2$/TiO$_2$ grown by co-deposition of Se and Ti.
As in TiSe$_2$/TiO$_2$, there is no band folding between $\Gamma$ and M, which indicates the suppression of CDW, and very large electron doping level of $n\approx 0.22$ electron/unit cell.
Also below the main band $\alpha$ near M there is a ``copy" band $\alpha'$, shown in Fig. 1(i), similar to the case of TiSe$_2$/TiO$_2$.

We further compared the band structure of TiSe$_2$/graphene and the TiSe$_2$ films grown on TiO$_2$.
Figure 2 (a-c) display the deeper valence bands for 1ML TiSe$_2$/graphene, 6ML TiSe$_2$/TiO$_2$ and Se:TiO$_2$, respectively.
In TiSe$_2$/TiO$_2$, there are two bands which share similar dispersion but are separated by 350 meV.
There is a hint of the band $\gamma'$ in Se:TiO$_2$ as well, but the signal is much weaker.
The origin of the band $\gamma'$ in TiSe$_2$ films on TiO$_2$ substrate needs further work to elucidate.
Besides this difference, the vast majority of band features are similar with the exception of a chemical potential shift in the cases of TiSe$_2$/TiO$_2$ and Se:TiO$_2$.

Figure 2 (d) shows the core level spectra of TiSe$_2$/graphene, 7ML TiSe$_2$/TiO$_2$ and Se:TiO$_2$.
The Se double peaks are similar for all three systems in both position and magnitude.
The Ti peaks for TiSe$_2$/TiO$_2$ and Se:TiO$_2$ are shifted towards lower binding energy and have slightly different shapes. 
Previous reports discuss the creation of Ti$^{3+}$ states on the surface of vacuum annealed TiO$_2$\cite{TiO23p}, which will cause additional shoulders with lower binding energy in the Ti $2p$ X-ray photoemission spectra.
Thus we may explain the energy shift in the core level spectra of TiSe$_2$/TiO$_2$ and Se:TiO$_2$ by the formation of Ti$^{3+}$ states.
However, this energy shift is also seen for 7ML TiSe$_2$/TiO$_2$, where the signal from the Ti$^{3+}$ ions in TiO$_2$ substrate should diminish.
With an estimated probing depth of 0.6 nm \cite{LINDAU1974}, the Ti$^{3+}$ signature from TiO$_2$ should diminish by a factor of 1,000 for a 7ML film compared to the bare TiO2 substrate.
This is in contradiction to the observation that Se:TiO$_2$ and TiSe$_2$/TiO$_2$ have similar Ti $2p$ photoemission spectra.
Such an observation suggests that the electron doping on TiSe$_2$/TiO$_2$ may induce Ti$^{3+}$ ions in TiSe$_2$ films as well.
The measurement of deeper valence band and core level further verifies that the films on TiO$_2$, grown by either Se deposition or Ti/Se co-deposition, are TiSe$_2$.

\section{Surface Characterization}
Next we continue to characterize the quality of the TiSe$_2$ films on TiO$_2$.
TiSe$_2$ is a layered material with hexagonal lattice and a lattice constant of $a=0.354 $~nm.
TiO$_2$ (100) surface is rectangular with lattice parameters $a = 0.459$ nm, $c = 0.295$ nm. 
We can view the (100) surface of the substrate as a distorted hexagonal lattice, but this still leads to a large and anisotropic strain of more than 20\%, as is shown in Figure 3(a).
Therefore, it seems unlikely to achieve layer-by-layer growth of TiSe$_2$/TiO$_2$. 
Nonetheless, Reflective High Energy Electron Diffraction (RHEED) and Scanning Tunneling Microscope (STM) topography show epitaxial growth of TiSe$_2$ for both TiSe$_2$/TiO$_2$ and Se:TiO$_2$.
The strain is $<$ 2\% determined by RHEED diffraction pattern. 
The RHEED images, oscillations of RHEED intensity and STM topography are shown in Figure 3(d-i).
Intriguingly, we can see RHEED intensity ``oscillations" for Se:TiO$_2$, even though the reaction is between deposited Se atoms and Ti atoms on the substrate.
These results, combined with the clarity in the ARPES spectra, demonstrate the excellent surface quality of the film for both Se deposition and co-deposition.

\section{Analysis}
With the chemical composition and surface quality verified for TiSe$_2$/TiO$_2$ and Se:TiO$_2$, we now turn back to the explanation of their electronic band structures.
The two main features of TiSe$_2$ films on TiO$_2$ are the large electron doping and the suppression of the CDW.

Currently the origin of the CDW in TiSe$_2$ is still under debate.
Some attribute it to a Peierls charge density wave, which is created by a spontaneous crystal distortion driven by the electron-phonon interaction\cite{Salvo76,Suzuki84SSC}.
Many others describe the low-temperature phase as excitonic condensation where electrons and holes combine into excitons which then Bose condense\cite{Wilson77SSC,Cercellier2007}.
In either case, it is expected that external electron doping will suppress the CDW.
For the case of a Peierls CDW, increasing carrier density will alter the Fermi surface nesting conditions and promote superconductivity, which competes with the CDW.
For an excitonic condensate, doping will break the balance between electrons and holes and hence restraining the condensation.
Previous researchers have doped TiSe$_2$ by Cu intercalation \cite{Morosan2006, ZhaoFeng07prl} and ionic liquid gating \cite{LiNato15nature}, and both methods lead to the destruction of the CDW and the formation of a superconductivity dome in the phase diagram.

The most unusual aspect of this system is the magnitude of its doping.
Changing post-growth annealing temperature of TiSe$_2$ can control Se vacancy defects, which may act as electron dopants\cite{PengMa15prb}.
By default we do post-growth annealing at the growth temperature, which is 200 $^\circ$C for TiSe$_2$/graphene, and 360-380 $^\circ$C for TiSe$_2$/TiO$_2$.
Therefore, we systematically compare the films of all 3 growth modes with different post-growth annealing temperatures to investigate its effect on the electronic properties.
Figure 4(e-f) shows the Fermi surface of TiSe$_2$/graphene after low and high temperature post-growth annealing.
The Fermi surface becomes significantly larger after a mere 50$^\circ$C increase in annealing temperature (from 200 $^\circ$C to 250 $^\circ$C), indicating that the high temperature post-growth annealing can create Se vacancies and induce electron doping.
Nevertheless, if we immediately quench TiSe2/TiO2 and Se:TiO2 films to temperatures below 200 $^\circ$C after growth, the Fermi surface volume is still much larger than the case of TiSe$_2$/graphene, which is clear if we compare Fig. 4(a, c, e).
When the doping level is higher than 0.15 electron/Ti atom, as shown in Fig 4 (a-d,f), the CDW order is destroyed. 
This shows that TiO$_2$ substrate, directly or indirectly, induces large doping level into TiSe$_2$.

It is plausible to assume the interfacial charge transfer from work function difference as the origin of the electron doping in TiSe$_2$ films on TiO$_2$.
The work function difference between TiSe$_2$\cite{LiuWei16SA} and TiO$_2$\cite{ImanishiYoshihiro07JPCC} is as large as $\sim$ 1.2 eV, which may enable direct interfacial transfer of electrons from TiO$_2$ substrate to TiSe$_2$ film.
However, the effects of interfacial charge transfer are strongly limited by the thickness of films due to electronic screening from the films.
Since there is no thickness dependence of the Fermi surface size up to 24ML (see Supplemental Fig. 1), the work function induced interfacial charge transfer cannot be the main reason for the electron doping for TiSe$_2$ films on TiO$_2$.

Below we consider an indirect effect from the TiO$_2$ substrate: the strong interfacial bonding between TiSe$_2$ and TiO$_2$ leads to a growth window with higher growth temperature. More Se vacancies are created during growth under this condition compared to the case of TiSe$_2$/graphene, resulting in electron doping and suppression of CDW in TiSe$_2$.

The ideal growth temperature for TiSe$_2$/TiO$_2$ and Se:TiO$_2$ is consistently higher than that for TiSe$_2$/graphene.
The ideal growth temperature for TiSe$_2$/TiO$_2$ and Se:TiO$_2$ is between 350-380 $^\circ$C, below which the growth yields amorphous films.
On the other hand, if we grow TiSe$_2$/graphene at any temperature above 250 $^\circ$C, there will be no formation of new RHEED streaks. 
This means that at this temperature, the Se re-evaporation rate is to high to allow the growth of TiSe$_2$ on graphene. 
This substrate-induced growth temperature difference between graphene and TiO$_2$ can be explained by the growth dynamics. 
Graphene provides much weaker interfacial interaction, therefore the ideal growth temperature is low.
In contrast, TiO$_2$ surface tends to form stronger bonds with TiSe$_2$ on the interface, thus a higher growth temperature is needed to provide sufficient diffusion rate necessary to form pristine films.

At the growth temperature of TiSe$_2$/TiO$_2$ and Se:TiO$_2$ (380 $^\circ$C), Se re-evaporation can be very strong, according to the results of TiSe$_2$/graphene from both Fig 4(f) and a previous STM study \cite{PengMa15prb}.
As a result, during the growth process, there is already strong Se re-evaporation, leading to Se deficiency and electron doping, even though we are growing in a Se-rich regime.
However, the heavily doped TiSe$_2$/TiO$_2$ and Se:TiO$_2$ films generally show superior ARPES spectrum than the TiSe2/graphene films with higher post-growth annealing temperatures.
The sharper ARPES is indicative of improved crystalline quality.  
Thus the enhanced substrate-film bonding yields significant selenium defects and therefore large electron doping, while maintaining superior crystallinity overall.

To summarize, we can explain the suppression of charge density wave order by the large electron doping, and explain the electron doping by Se vacancies created by substrate induced change of growth conditions.
The doping is because of the change of growth condition instead of a direct effect of substrate, which can explain why this effect is independent of film thickness up to 24ML.

\section{Conclusion}
We grew TiSe$_2$ on rutile TiO$_2$ (100) substrates using MBE, in which substrate effects provide large doping to TiSe$_2$ that suppresses CDW order.
This demonstrates that, in the course of tuning the chemical potential of thin film samples, we can add a large and flexible offset by selecting and engineering the substrate.
The substrate-induced doping, compared to conventional methods including alkali metal atom intercalation and ionic liquid gating, provides a cleaner way of doping thin films in that it does not provide external impurities to the system.
In addition, our results demonstrate that pristine TiSe$_2$ thin films can form on reconstructed TiO$_2$ substrate after Se deposition, which may pave the way to new growth methods of other chalcogenide/oxide heterostructure, and improve our understanding of the chemical property of the reconstructed surface of oxides. 

Previous studies doped electrons into TiSe$_2$ to a maximal level of $\sim$0.06 electron per unit cell, and observed superconductivity\cite{Morosan2006,ZhaoFeng07prl,QianHasan2007prl}. 
In contrast, TiSe$_2$ films on TiO$_2$ reported in this work has a doping of $>$0.16 electron per unit cell without any external intercalation or liquid gating, entering into an unexplored regime in the TiSe$_2$ phase diagram. 
It would be interesting to do further experiments to investigate into the transport properties of TiSe$_2$ films on TiO$_2$.
We also plan to apply similar substrate engineering methods to other heterostructures of correlated materials as a powerful way to explore further in the phase diagrams.
\section*{Methods}
\subsection*{Growth}
For TiSe$_2$/TiO$_2$, Shinkosha STEP TiO$_2$ (100) substrates were mounted on a molybdenum sample holder with silver paste and placed into an MBE chamber. The base pressure of the MBE chamber was 8 $\times 10^{-11}$ torr.
Substrates were then degassed at 450 - 650 $^\circ$C, and cooled down to 360 $^\circ$C for growth.
Ultra-high purity titanium (99.995\%) and selenium (99.999\%) were then deposited onto the substrate.
The crystallinity was examined using reflection high energy electron diffraction (RHEED).
The films were then annealed at 380$^\circ$C for 2 hours (except for the ones noted with ``Low Temperature Anneal" in Fig. 4).
The growth conditions of Se:TiO$_2$ is similar to that of TiSe$_2$/TiO$_2$, except that before growth the substrates are  annealed at 810 - 860 $^\circ$C for 30 minutes in addition to degassing, and that only Se cell is used during the growth.

For TiSe$_2$/graphene, the TiSe$_2$ films were grown by MBE on bilayer graphene (BLG) epitaxially grown on 6H-SiC\cite{BLG}. The growth temperature is at around 200 $^\circ$C. The films were then annealed at growth temperature $^\circ$C for 2 hours (except for the ones noted with ``High Temperature Anneal" in Fig. 4).

The successful growth of Se:TiO$_2$ relies on the preparation of substrate.
TiO$_2$ rutile (100) surface undergoes a (3 $\times$ 1) reconstruction at temperatures above 670 $^\circ$C, exposing Ti atoms from surface layers to vacuum \cite{DieboldTiO2}. 
Only when we pre-anneal the substrate to the temperature range within 810 - 860 $^\circ$C will we grow epitaxial films with clear ARPES spectra.
If pre-annealing temperature is too low, the films will not form, possibly because there are not enough Ti atoms exposed to vacuum.
If pre-annealing temperature is above 860 $^\circ$C, the films will show ARPES spectra with very broad bands and scattered intensity, resulting from a very rough substrate surface.

On a side note, we want to alert chalcogenide thin film growers that substrate-film reaction and the resulting Se:TiO$_2$ may accidentally occur during the process of preparing other chalcogenide/titanate heterostructures, as the condition for the formation of Se:TiO$_2$ is in close proximity with the growth condition of many other chalcogenide films. 
For example, Se:TiO$_2$ can act as an impurity phase for the growth of FeSe/TiO$_2$ if the substrate is preannealed at temperatures above 790 $^\circ$C, giving hexagonal lattice with Fermi surfaces and electronic structures very similar to what is reported in this work.
Similar mechanisms of TiSe$_2$ formation may also happen for chalcogenide films on other titanate substrates including SrTiO$_3$ and anatase TiO$_2$.

\subsection*{Measurement}
After growth, the films were transferred \textit{in situ} to the ARPES end station of the Stanford Synchrotron Radiation Lightsource beamline 5-2. 
The base pressure in the ARPES chamber was better than $4\times10^{-11}$ torr.
Circular right polarization was used during the measurement. The high symmetry cuts were done with photon energy of 46 eV, and the energy resolution is better than 25 meV; Reciprocal space maps were measured using photon energy of 82 eV, and the energy resolution is better than 44 meV. The angular resolution is better than 0.4$^\circ$.

\pagebreak
\begin{figure*}
\centering
\includegraphics[]{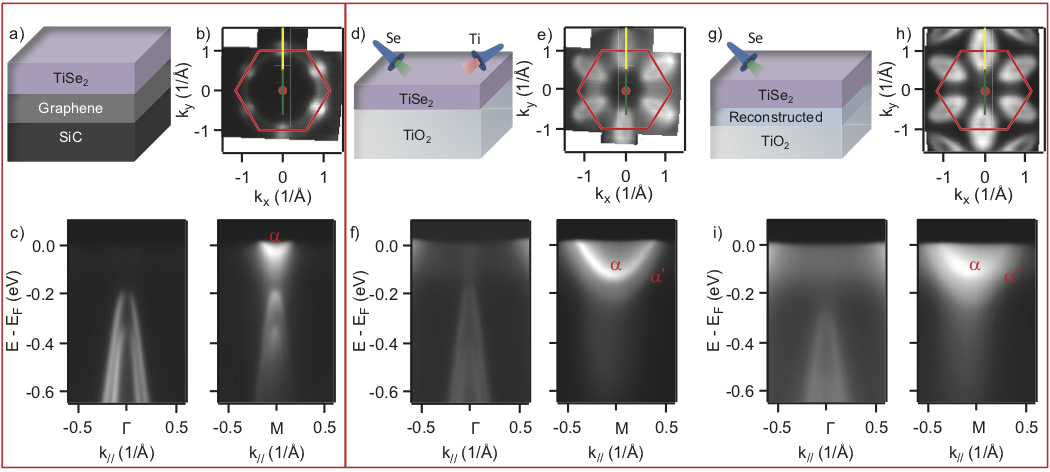}
\caption{\label{fig:1}(a) sketch of the heterostructures for TiSe$_2$/graphene. (b) photoemission intensity maps near $E = E_F$ for an energy window of 10 meV for 1ML TiSe$_2$/graphene. The yellow and green lines represent the direction of high-symmetry scans through  zone center and zone boundary, respectively. (c) band structure for 1ML TiSe$_2$/graphene along the high symmetry scans across the zone center (green line in (b, e, h)) and the zone boundary (yellow line in (b, e, h)).
(d, e, f) same as (a, b, c), except that the sample for (e) is 15ML TiSe$_2$/TiO$_2$ and the sample for (f) is 10.5ML TiSe$_2$/TiO$_2$.
(g, h, i) the same as (a, b, c), except that the samples are Se:TiO$_2$.
All data were taken at 20 K. (b, e, h) taken with 82 eV photons, (c, f, i) taken with 46 eV photons.
%The spectra of TiSe$_2$/TiO$_2$ are largely layer independent; please see Supplemental Information for details.
%(a) sketch of the heterostructures for TiSe<sub>2</sub>/graphene. (b) photoemission intensity maps near E = E<sub>F</sub> for an energy window of 10 meV for 1ML TiSe<sub>2</sub>/graphene The yellow and green lines represent the direction of high-symmetry scans through zone center and zone boundary, respectively. (c) band structure for 1ML TiSe<sub>2</sub>/graphene along the high symmetry scans across the zone center (green line in (b, e, h)) and the zone boundary (yellow line in (b, e, h)). (d, e, f) same as (a, b, c), except that the sample for (e) is 15 ML TiSe<sub>2</sub>/TiO<sub>2</sub> and the sample for (f) is 10.5 ML TiSe<sub>2</sub>/TiO<sub>2</sub>. (g, h, i) the same as (a, b, c), except that the sample is Se:TiO<sub>2</sub>. All data were taken at 20 K. (b, e, h) taken with 82 eV photons, (c, f, i) taken with 46 eV photons.
}
\end{figure*}

\begin{figure*}
\centering
\includegraphics[]{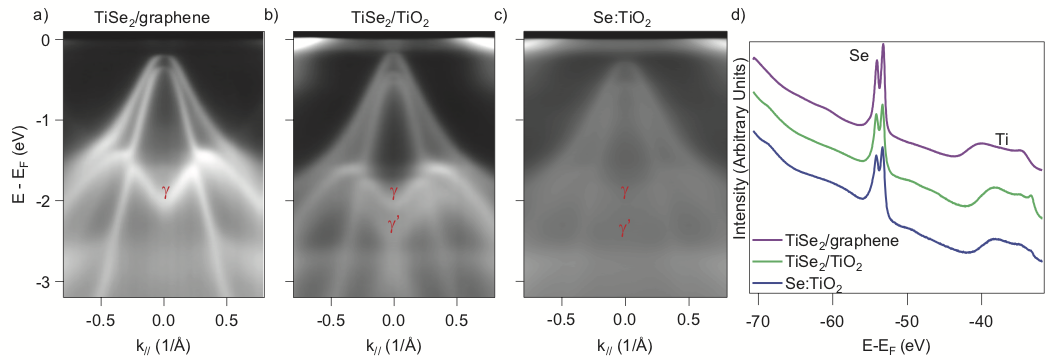}
\caption{\label{fig:2}(a) The band structure for 1ML TiSe$_2$/graphene at the zone center (see the green line in Fig 1(b)); (b) The band structure for 10.5ML TiSe$_2$/TiO$_2$ at the zone center; (c) The band structure for Se:TiO$_2$ at zone center; (d) The core level photoemission intensity of 1ML TiSe$_2$/graphene, 7ML TiSe$_2$/TiO$_2$ and Se:TiO$_2$.
%(a) The band structure for 1ML TiSe<sub>2</sub>/graphene at the zone center (see the green line in Fig 1(b)); (b) The band structure for 10.5ML TiSe<sub>2</sub>/TiO<sub>2</sub> at the zone center; (c) The band structure for Se:TiO<sub>2</sub> at zone center; (d) The core level photoemission intensity of 1ML TiSe<sub>2</sub>/graphene, 7ML TiSe<sub>2</sub>/TiO<sub>2</sub> and Se:TiO<sub>2</sub>.
All data were taken at 20 K. (a-c) taken with 46 eV photons, (d) taken with 82 eV photons.
}
\end{figure*}

\begin{figure*}
\centering
\includegraphics[]{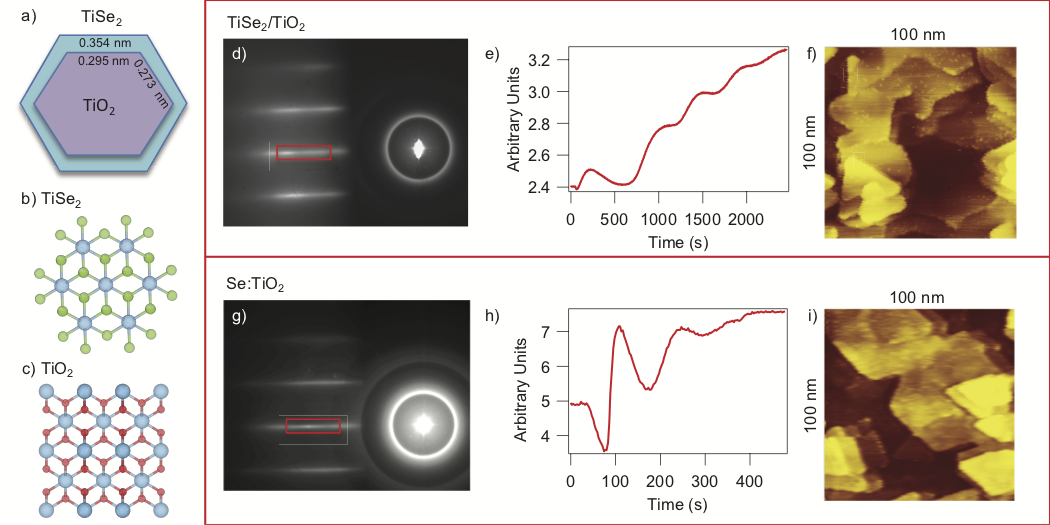}
\caption{\label{fig:3}(a) Comparison between the lattices of TiSe$_2$ and TiO$_2$; (b) The (0001) projection of the atomic structure of TiSe$_2$, where blue spheres indicate Ti atoms, and green spheres indicate Se atoms; (c) The (100) projection of the atomic structure of TiO$_2$, where blue spheres indicate Ti atoms, and red balls indicate Oxygen atoms; (d-f) Surface characterization of TiSe$_2$/TiO$_2$ films: (d) RHEED pattern for a 21ML sample; (e) The evolution of RHEED intensity integrated from the red rectangle versus time during growth for a 15ML sample; (f) the STM topography measurement for the same sample as (e); (g-i) The same as (d-f), but for Se:TiO$_2$ film.
%(a) Comparison between the lattices of TiSe<sub>2</sub> and TiO<sub>2</sub>; (b) The (0001) projection of the atomic structure of TiSe<sub>2</sub>, where blue spheres indicate Ti atoms, and green spheres indicate Se atoms; (c) The (100) projection of the atomic structure of TiO<sub>2</sub>, where blue spheres indicate Ti atoms, and red balls indicate Oxygen atoms; (d-f) Surface characterization of a TiSe<sub>2</sub>/TiO<sub>2</sub> film: (d) RHEED pattern for a 21ML sample; (e) The evolution of RHEED intensity integrated from the red rectangle versus time during growth for a 15ML sample; (f) the STM topography measurement for the same sample as (e); (g-i) The same as (d-f), but for Se:TiO<sub>2</sub> film.
}
\end{figure*}

\begin{figure*}
\centering
\includegraphics[]{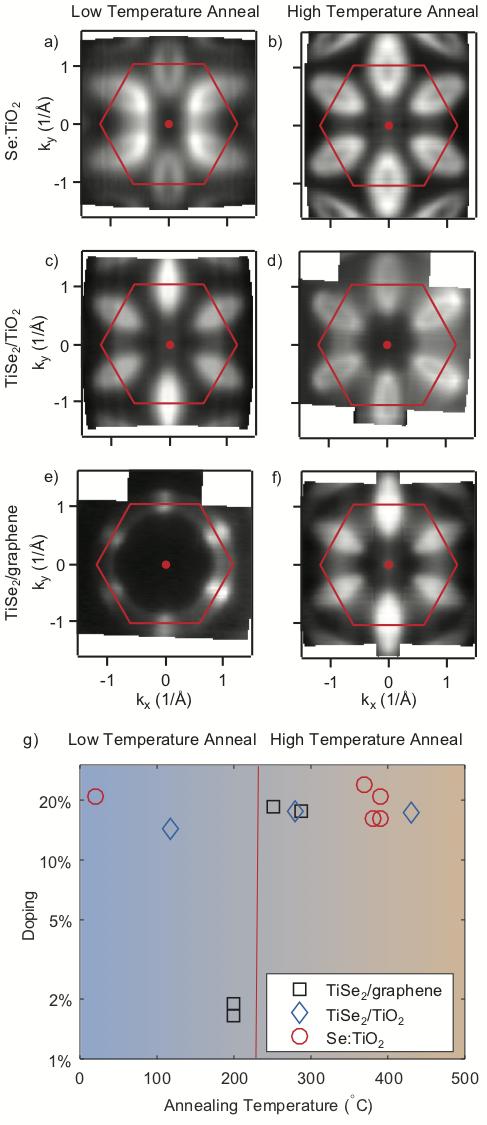}
\caption{\label{fig:4}(a-f) ARPES intensity mapping near $E = E_F$ for an energy window of 10 meV, measured at 20 K. The Brilloin zone of bulk TiSe$_2$ is plotted in red for comparison. (a) Se:TiO$_2$, quenched to 30 $^\circ$C immediately after growth; (b) Se:TiO$_2$, annealed at 390 $^\circ$C; (c) TiSe$_2$/TiO$_2$,9ML, quenched to 120 $^\circ$C; (d) TiSe$_2$/TiO2, 15ML, annealed at 280 $^\circ$C; (e) TiSe$_2$/graphene, 1ML, annealed at 220 $^\circ$C; (f) TiSe$_2$/graphene, 1ML, annealed at 280 $^\circ$C. (g) The relationship between doping level and annealing temperature for Se:TiO$_2$, TiSe$_2$/TiO$_2$ and TiSe$_2$/graphene. The red line is a guide of the eye indicating the separation between “Low Temperature Anneal” and “High Temperature Anneal”. All data taken with 82 eV photons.
%(a-f) ARPES intensity mapping near E = E<sub>F</sub> for an energy window of 10 meV, measured at 20 K. The Brilloin zone of bulk TiSe<sub>2</sub> is plotted in red for comparison. (a) Se:TiO<sub>2</sub>, quenched to 30 °C immediately after growth; (b) Se:TiO<sub>2</sub>, annealed at 390 °C; (c) TiSe<sub>2</sub>/TiO<sub>2</sub>,9 ML, annealed at 120 °C; (d) TiSe<sub>2</sub>/TiO2, 15 ML, annealed at 280 °C; (e) TiSe<sub>2</sub>/graphene, 1 ML, annealed at 220 °C; (f) TiSe<sub>2</sub>/graphene, 1 ML, annealed at 280 °C. (g) The relationship between doping level and annealing temperature for Se:TiO<sub>2</sub>, TiSe<sub>2</sub>/TiO<sub>2</sub> and TiSe<sub>2</sub>/graphene. The red line is a guide of the eye indicating the separation between “Low Temperature Anneal” and “High Temperature Anneal”. All data taken with 82 eV photons.
}
\end{figure*}

\pagebreak

\begin{acknowledgement}
We would like to thank B. Moritz, C.-J. Jia, Y. He, and S.-D. Chen for valuable discussions. 
This work is supported by the Department of Energy,  Office of Science, Basic Energy Sciences, Materials Sciences and Engineering Division, under Contract DE-AC02-76SF00515.
Use of the Stanford Synchrotron Radiation Lightsource, SLAC National Accelerator Laboratory, is supported by the U.S. Department of Energy, Office of Science, Office of Basic Energy Sciences, also under Contract No. DE-AC02-76SF00515.
\end{acknowledgement}
\bibliography{achemso-demo}

\end{document}